\documentclass{article}

\PassOptionsToPackage{numbers, compress}{natbib}
\usepackage[final]{neurips_2018}

\usepackage{setspace} 
\usepackage[utf8]{inputenc} 
\usepackage[T1]{fontenc}    
\usepackage{hyperref}       
\usepackage{url}            
\usepackage{booktabs}       
\usepackage{amsfonts}       
\usepackage{nicefrac}       
\usepackage{microtype}      
\usepackage{graphicx}
\makeatletter
\newcommand{\printfnsymbol}[1]{%
  \textsuperscript{\@fnsymbol{#1}}%
}
\makeatother
\title{Emotion Recognition from Speech}
\author{%
  Kannan Venkataramanan\printfnsymbol{1}  \\
  \texttt{kv942@nyu.edu}
  \And
  Haresh Rengaraj Rajamohan\thanks{Equal contribution.} \\
  \texttt{hrr288@nyu.edu} \\
}

\begin{document}

\maketitle

\begin{abstract}
  In this work, we conduct an extensive comparison of various approaches to speech based emotion recognition systems. The analyses were carried out on audio recordings from Ryerson Audio-Visual Database of Emotional Speech and Song (RAVDESS). After pre-processing the raw audio files, features such as Log-Mel Spectrogram, Mel-Frequency Cepstral Coefficients (MFCCs), pitch and energy were considered. The significance of these features for emotion classification was compared by applying methods such as Long Short Term Memory (LSTM), Convolutional Neural Networks (CNNs), Hidden Markov Models (HMMs) and Deep Neural Networks (DNNs). On the 14-class (2 genders $\times$ 7 emotions) classification task, an accuracy of 68\%  was achieved with a 4-layer 2 dimensional CNN using the Log-Mel Spectrogram features. We also observe that, in  emotion recognition, the choice of audio features impacts the results much more than the model complexity.   
\end{abstract}

\footnotetext{Please refer to \url{https://github.com/rajamohanharesh/Emotion-Recognition} for the code.}
\section{Introduction}

According to a report from the United Nations \cite{UN}, an increasing number of people will interact with a voice assistance machine than with their partners in the next five years. With proliferation of Virtual Personal Assistants (VPA) such as Siri, Alexa and Google Assistant in our day-to-day interactions, they fill a role of answering our questions and fulfilling our requests quickly and accurately. Though these assistants understand our commands, they are not proficient enough in recognizing our mood and reacting accordingly. Therefore, it is pertinent to develop an efficient emotion recognition system which can enhance the capabilities of these assistants and revolutionize the whole industry.

Speech is a rich, dense form of communication that can convey information effectively. It contains two types of information, namely linguistic and paralinguistic. The former refers to the verbal content, the underlying language code, while the latter refers to the implicit information such as body language, gestures, facial expressions, tone, pitch, emotion etc. Para linguistic characteristics can help understand the mental state of the person (emotion), gender, attitude, dialect, and more \cite{Yamashita}. Recorded speech has key features that can be leveraged to extract information, such as emotion, in a structured way. To get such information would be invaluable in facilitating more natural conversations between the virtual assistant and the user since emotion color everyday human interactions. 

There are two widely used representations of emotion: continuous and discrete. In the continuous representation, the emotion of an utterance can be expressed as continuous values along multiple psychological dimensions. According to Ayadi, Kamel, \& Karray (2011) \cite{Ayadi}, “emotion can be characterized in two dimensions: activation and valence.” Activation is the “amount of energy required to express a certain emotion” (p. 573) and research has shown that joy, anger, and fear can be linked to high energy and pitch in speech, whereas sadness can be linked to low energy and slow speech. Valence gives more nuance and helps distinguish between emotions like being angry and happy since increased activation can indicate both (p. 573). In the discrete representation, emotions can be discretely expressed as specific categories, such as angry, sad, happy, etc.

The performance of an emotion recognition system purely relies on features/representation extracted from the audio. They are broadly classified into time-based and frequency-based features. Extensive research has been carried out to weigh in the pros and cons of these features. There is no one particular sound feature which can perform well across all the sound signal processing tasks. Additionally, features are hand-crafted to suit the requirements of the problem in hand. With the advent of deep learning techniques, we have been successful in extracting the hierarchical representation of the speech from these features and identifying the underlying emotion in the speech. Hence, the performance of the model in a particular speech recognition task is much more dependent on the choice of the feature than the model architecture. 

Various experiments have been carried out in the past to identify emotions from speech for different languages and accents.Chenchah and Lachiri \cite{Chenchah} studied the performance of Mel-Frequency Cepstral Co-efficients and Linear Frequency Cepstral Coefficient (LPCC) in identifying the emotions using Hidden Markov Model (HMM) and Support Vector Machines (SVM). The developed model yielded a 61\% accuracy on Surrey Audio-Visual Expressed Emotion (SAVEE) Database. Parthasarathy and Tashev \citep{Srinivas} have compared DNNs, RNNs and 1D-CNN models on MFCC features from a Chinese language dataset. They' ve achieved 56\% accuracy with 1D CNN model.

This study focuses on identifying the best audio feature and model architecture for emotion recognition in speech. The experiments were carried out on "The Ryerson Audio-Visual Database of Emotional Speech and Song (RAVDESS)" dataset \citep{Livingstone}. The robustness of the model was assessed by predicting the emotions of speech utterances on a completely different dataset, the "Toronto Emotional Speech Set (TESS)" dataset. A four layer 2D-CNN architecture with Log-Mel Spectogram audio features yielded the maximum of accuracy of ~70\% on the test set and 62\% on TESS dataset.

\section{Datasets}
\subsection{Data Selection}

There are three main components to designing a SER: choosing an emotional speech database, feature selection from audio data, and the classifiers to detect emotion \citep[p. 573]{Ayadi}. Ryerson Audio-Visual Database of Emotional Speech and Song (RAVDESS) dataset is a validated multi-modal database of emotional speech and song. This gender-balanced database “consists of 24 professional actors, each performing 104 unique vocalizations with emotions that include: happy, sad, angry, fearful, surprise, disgust, calm, and neutral” \citep[p.2]{Livingstone}. Each actor enacted 2 statements for each emotion: “Kids are talking by the door” and “Dogs are sitting by the door.” These statements were also recorded in two different emotional intensities, normal and strong, for each emotion, except for neutral (normal only) (p. 2-3). Actors repeated each vocalization twice (p. 11). There are a total of 1440 speech utterances and 1012 song utterances.

The RAVDESS dataset is very rich in nature given that it does not suffer from gender bias, consists of wide range of emotions and at different level of emotional intensity. Other datasets, such as Surrey Audio-Visual Expressed Emotion \citep{Haq} (SAVEE) and Toronto Emotional Speech Set \citep{Dupuis} (TESS), consisted of audios from only male and female actors respectively. We also observe that the RAVDESS dataset is equally distributed across all emotion classes ($\sim$ 15\%), so it does not suffer from any class-imbalance problems.

Additionally, extensive validation and reliability tests have been performed by the creators of RAVDESS dataset. From a “pseudo-randomly chosen set of 298 stimuli, consisting of 174 speech and 124 song presentations,” 247 naive participants were asked to make three judgements on three classes: “category of the emotion, strength of the emotion, and genuineness of the emotion” \citep[p. 12]{Livingstone}.

From Figure 1, we observe that approximately 73\% of the rater chosen emotion were well-acted by the actors, ensuring the reliability of the classification of the emotions and the audio content. Additionally, we also observe that human raters found it difficult to distinguish neutral \& calm emotions. We manually listened to the audio files from RAVDESS and also felt that emotions calm and neutral sounded very similar to each other. Hence, we decided to merge these two emotions into a single class. The data distribution across gender and emotion classes is shown in Figure 2.

\begin{center}
\begin{figure}[htp]
    \centering
    \includegraphics[width=11cm]{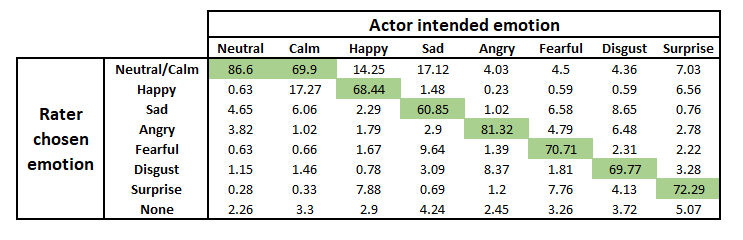}
    \caption{Dataset Validation (Livingstone \& Russo, 2018)}
    \label{fig:validation}
\end{figure}
\end{center}

\begin{center}
\begin{figure}[htp]
    \centering
    \includegraphics[width=13.5cm]{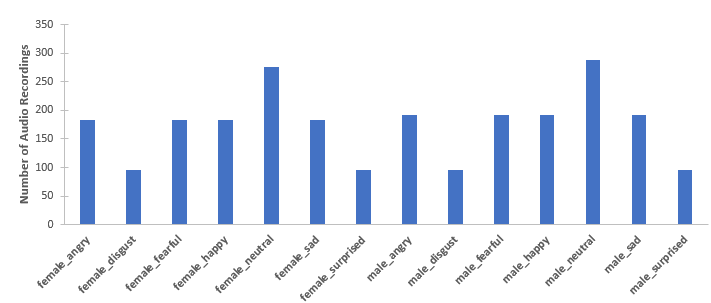}
    \caption{Data Distribution across gender and emotion}
    \label{fig:validation}
\end{figure}
\end{center}

\subsection{Selection Bias \& Limitations}
There are shortcomings in the dataset. Firstly, the way an emotion is exhibited is strongly rooted on the language, accent, dialect and cultural background. A model trained to identify emotion on English dataset might not be able to identify emotions in Chinese/Indic language speech utterances. RAVDESS dataset naturally suffers selection bias because the dataset was created using 24 English-speaking actors from the Toronto, Canada, who exhibit strong North-American characteristics. Secondly, the dataset was created using trained actors rather than using natural instances of emotions. Due to this limitation, the model should be extensively validated before deployment. A final limitation was the inclusion of only two statements, limiting the lexical variability of the database. As such, the onus of understanding the emotional patterns and not the words itself fall heavily on feature engineering and modeling. Diversity in the statements could have reduced the challenges for us.

\subsection{Data Preparation}

Each audio file contains a 7-part numerical identifier each denoting the modality, vocal channel, emotion, emotional intensity, statement, repetition and the actor respectively. The naming convention followed a pattern, wherein odd actors and even actors denoted male and female sex respectively. We extracted all these information from the file names into metadata. The target variable is the emotion that the audio recording was classified as.

\subsection{Data Cleaning}
Each recording was approximately 3 seconds long. Audio recordings were trimmed to remove silences both at the start and at the end. Though the audios were professionally recorded, there existed minute noise patterns in the data. We tried various signal processing techniques, such as filtering and voice-activity detection (VAD) to weed out the noise. Spectral subtraction \cite{Fux} is a well known noise reduction technique which aims to remove the background noise (i.e., an additive noise) by subtracting an estimation of the noise spectrum from the noisy speech spectrum. We implement Wiener filtering \cite{Weiner} to filter out the noise from the corrupted signal and provide an clear version of the underlying signal. Post-implementation, we observed a significant improvement in the quality of the dataset without compromising any important audio content.

\begin{center}
\begin{figure}[htp]
    \centering
    \includegraphics[width=13.5cm]{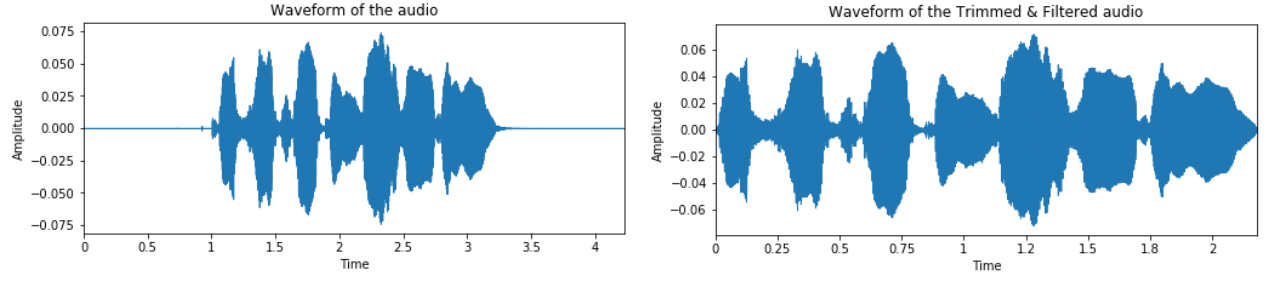}
    \caption{Waveform of pure audio pre and post data cleaning}
    \label{fig:validation}
\end{figure}
\end{center}

\section{Feature Engineering and Modelling}
\label{gen_inst}

\subsection{Feature Selection}

Audio features can be broadly classified into two categories, namely time-domain features and frequency-domain features. Time-domain features include the short-term energy of signal, zero crossing rate, maximum amplitude, minimum energy, entropy of energy. These features are very easy to extract and provide a simpler way to analyze audio signals. Under limited data, frequency domain features reveal deeper patterns in the audio signal, which can potentially help us identify the underlying emotion of the signal. Frequency-domain features include spectograms, Mel-Frequency Cepstral Coefficients (MFCCs), spectral centroid, spectral rolloff, spectral entropy and chroma coefficients \cite{Tomas}.

During Exploratory Data Analysis, an extensive analysis of each feature was carried out. However, for purpose of this report, we restricted ourselves to two main features, namely Mel-Frequency Cepstral Coefficients and Mel-spectograms. Refer appendix (Section 7.3,7.4) for additional information on feature selection.

\subsubsection{Mel-Frequency Cepstrum (MFC)}

Mel-Frequency Cepstrum is a representation of the short-term power spectrum of a sound by transforming the audio signal through a series of steps to mimic the human cochlea. The Mel scale is important because it better approximates human-based perception of sound as opposed to linear scales \cite{Shaw}. In filter-source theory, "the source is the vocal cords and the filter represents the vocal tract." The length and shape of the vocal tract determine how sound is outputted from a human and the cepstrum ("spectrum of the log of the spectrum") can describe the filter, i.e., represent sound in a structured manner \citep[p.7]{Tomas}. Mel-Frequency Cepstral Coefficients (MFCC) are coefficients which capture the envelope of the short time power spectrum. Refer \cite{Melspec} for a detailed explanation.

MFCCs are computed as follows:

{\footnotesize
\begin{itemize}
  \item The audio signal is cut into short frames to ensure the stationarity of the audio signal.
  \item For each frame, a periodogram, which identifies the frequencies present in the frame, is estimated.
  \item A Mel filter bank, which merges periodogram bins by summing up the energy, is performed to get an idea of how much energy exists in various frequency regions. This step is performed to resonate with the way human cochlea works.
  \item Since human perception of sound does not follow a linear scale, the spectral energies are transformed into log scale to transform the non-linear frequency pattern to a linear scale, facilitating direct inference.
  \item Generally these filterbanks overlap with each other resulting in energies across frequency band being correlated with each other. A discrete cosine transform (DCT) decorrelates these energies. This results in different coefficients which can be used for further analysis.
\end{itemize}
}
The higher MFC coefficients represent fast changes in the filterbank energies. Hence, the number of MFCCs is also a variable which was iterated in this experiment. Additionally, MFCC features describe only the power spectral envelope of a single frame. A speech recognition system could benefit from the rate of change/trajectories of these MFCC features. These are called delta coefficients of MFCC. Similarly, rate of change of trajectories (acceleration), i.e. delta-delta coefficients also provide insights for the Automatic Speech Recognition (ASR) system. These coefficients can be stacked along with MFCCs for the modelling purpose. In this study, various models was trained using the three configurations: MFCC, MFCC + Delta  and MFCC + Delta-Delta coefficients. Refer appendix (Section 6.2) for the plots of MFCC, their deltas and delta-deltas.

\subsubsection{Mel-Spectograms}
A spectogram is a time vs frequency representation of an audio signal. Different emotions exhibit different patterns in the energy spectrum. Mel-spectogram is a representation of the audio signal on a Mel-scale. The logarithmic form of mel-spectogram helps understand emotions better because humans perceive sound in logarithmic scale. Therefore, the log-mel spectogram corresponds to the time vs. log-mel frequency representation, which was obtained in step 4 during MFCC computation. Refer appendix (Section 7.5) for the plots of log-mel spectrogram of different emotions in the dataset.

We observe that MFCC features and Log-Mel Spectograms can be represented as images, and these images can be fed to deep learning techniques such as CNN, RNN networks to classify the emotion of an audio.

\subsection{Models}
\subsubsection{Convolutional Neural Networks}
The tremendous strides made in the recent years in image recognition tasks is in large part due to the advent of Convolutional Neural Networks \citep{Krizhevsky} (CNNs). CNNs are good at automatically learning relevant features from high dimensional input images. CNNs use shared kernels (weights) to exploit the 2D correlated structure of image data. Max-pooling is added to CNNs to introduce invariance wherein only the relevant high dimensional features are learned for various tasks like classification, segmentation etc. Surprisingly, CNNs are good at audio recognition tasks. This is because, when Log mel filter bank is applied to FFT representation of raw audio , a "linearity" is introduced in the log-frequency axis allowing us to perform convolution operation along the frequency axis. Else we would have to use different filters(or kernels) for different frequency ranges. This property along with CNN's superior representation power helps the model learn the underlying patterns effectively from short time-frames resulting in state of the art performance in Speech based Emotion Recognition systems.

3D CNNs \citep{Ji} tend to perform well in video understanding tasks. In video data, time correlation is present between the frames but not within each frame. But in the case of reshaped 3D-audio, there will be time-dependency along two axes- small time lag along one axis and longer time lag along the other. So, we wanted to see if modelling both short term dependency and long term dependency in the data improves our SER system. Thus, the 3D CNN model was also considered in our work. The length of each frame in the 3D audio was set to be 250 ms, which was found to be the smallest optimal length for inferring emotions in \citep{Srinivas}.

Additionally, multi layer 1D CNNs were also trained on raw audio to identify patterns in the sound waveform. 

\subsubsection{Recurrent Neural Networks}
Recurrent neural networks \citep{Hochreiter} are widely used in time-series problems. In RNNs, at each time-step, a hidden state is computed from the input and this hidden state along with next input is used to compute the hidden state at the next time step. So RNNs maintain a memory which is used to understand the larger context for prediction. LSTM is a class of RNN,  which efficiently deals with the exploding and vanishing gradient problems, usually encountered in vanilla-RNNs. As explained earlier we assume stationarity in small time frames (10 ms) and the features extracted in this time frame will be the input at each timestep in RNN. To actually understand emotion, we need larger time-steps(~250ms) , which means the RNN needs to have longer memory for good performance. So, LSTM is a natural candidate for building SER systems. 

Similar to the idea of 3D-CNNs, 2D CNN-LSTM (frame length 250 ms) \citep{Jeff} models was also implemented for SER in our work. Additionally, 1D CNN-LSTM models were trained on raw audio files to capture the temporal emotional patterns in the raw-audio signal.

\subsubsection{Hidden Markov Models}
Before RNNs became popular, HMMs \citep{Rabiner} were widely used for speech recognition tasks like emotion recognition \citep{Schuller}. HMMs work under the assumption observations come from some number of hidden states and the transitions between these states satisfy Markov property. HMMs are generative models. One serious limitation of HMM is that they perform poorly when the data being modeled is non-linear. We implemented 2 types of HMMs - Gaussian HMMs and GMM HMMs, where the emission characteristics are Gaussian and Gaussian mixtures respectively. The hidden states in these HMMs are the phones (smallest discrete sound) and we classify based on the maximum likelihood of the sequence of phones coming from different classes.

During training, one HMM per class is trained with data from the respective class with different number of hidden states (Number of HMMs is equal to the number of emotion classes). In testing, the likelihood probability $p(D|c)$ (D is the utterance, c is the class) of the utterance given class is calculated for all the classes and the $label\,=\,argmax_{c}p(D|c)$.

\section{Experiments}
\subsection{Data}
MFCCs were extracted with a window size of 10 ms and hop length of 5 ms. The number of MFCCs per frame was also tuned in our analysis and it was found that there was no performance gain beyond 25 features per frame. Apart from this, features like pitch, magnitude and mean squared energy, their deltas and delta-deltas were also added to the MFCCs. 128 Log-Mel spectrogram  features were extracted from input audios with a window size and hop length of 0.014 sec and 0.0035 sec respectively.  Refer to section 2 \& 3 for detailed explanation.

\subsection{Train, Validation and Test Data Creation}

Emotion recognition experiments are broadly categorized into speaker-dependent and speaker-independent experiments. Speaker-dependent experiments contain audio instances of different emotions of the same actor in the the training, validation and test datasets. This means the dataset is randomly split and the training data could over-fit to one particular actor, leading to bias of the model. Hence, randomly splitting is one form of data leakage for this task and is not advisable. On the other hand, speaker-independent experiments are  experiments where the training, validation and test data consist of audio instances from different actors. Speaker-independent training ensures that the models are robust and will be able to identify emotions irrespective of the actor. Hence, audio instances of actors 1-20, actors 21 \& 22 and actors 23 \& 24 were used for training, validation and testing respectively. Since, both validation and testing datasets contain instances of both male (odd actors) and female (even actors) actors, we have ensured that the models are not tuned to a particular gender. Also, by splitting it this way, we have also ensured that there are no actor-characteristic leakage in the dataset. 

To ascertain that the model is robust enough, a stringent assessment of the model was performed on a completely different dataset, TESS. Since the origin of TESS is unrelated to RAVDESS, but the recordings are from North-American speakers, we strongly believe that if the model could predict the emotions accurately in TESS, we can confidently deploy the model in action. Hence the data cleaning steps were also performed on the TESS dataset.

\subsection{Methodology}
Given the data is almost equally distributed, accuracy is a valid metric to compare the performance of the models. So the model selection metric was  chosen to be vanilla unweighted accuracy. For training models with fixed size inputs, all the input audios were made into 3 seconds by trimming or padding them appropriately. 

All the models were trained for 100 epochs with different batch sizes (depending on complexity of architecture). The models were initially trained with stochastic gradient descent \citep{Bottou} (SGD) optimizer. Due to the slower convergence with SGD, ADAM \citep{Diederik} optimizer was used for the later experiments with default parameters.  Models were saved by monitoring the accuracy on the validation set.

We trained a baseline DNN architecture with MFCC features (claimed to be the best suited for emotion recognition by practitioners) on the 7 class prediction and found that the network was unable to understand recordings from the "Surprised" class. We felt that this could be due to the shortage of "Surprised" class in the data. Hence, "Surprised" class was removed from the dataset and further analyses were conducted only using the 6 classes - angry, sad, neutral, disgust, happy and fearful. Increasing the number of MFCCs and adding deltas gave us the best DNN performance. 

As seen in the visualization of MFCCs and Log-Mel spectrogram, MFCCs have very little information in the higher frequencies. So, we expect log-mel spectrogram features to give better results in this study.

1D CNNs and 2D CNNs were implemented on 6 emotion class classification and 12 gender+emotion class classification parallelly and it was observed that including gender gave better performance. Since CNNs are natural feature extractors, 1D CNN and 1D CNN-LSTM architectures were also trained on the raw audio input. 

2D CNNs were implemented on the engineered features such as MFCCs and Log-mel spectrogram. The training of 2D CNNs started with 2 convolutional layers with $3\times3$ filters and max pooling with $2\times2$ filters with stride 2. They were tuned by adding more convolutional layers and increasing the filter sizes in the initial layers. It was found that increasing the depth beyond 4 layers did not improve performance. During the later experiments, the "Surprised" class was added to our analysis. 

\begin{figure}[htp]
    \centering
    \includegraphics[width=13.5cm]{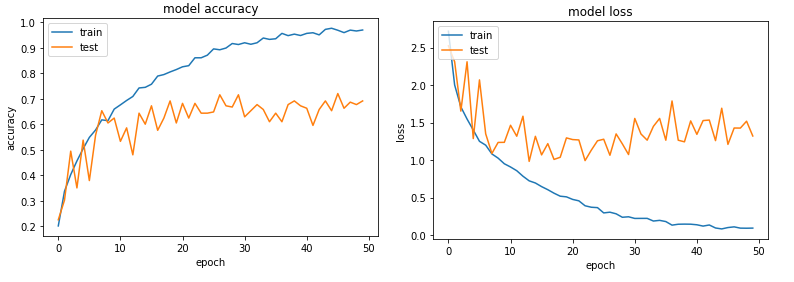}
    \caption{Training Accuracy and Loss}
    \label{fig:validation}
\end{figure}

The final pooling layers were assigned a higher filter size and stride of 4. This helped in reducing the number of parameters in the fully connected layer, when the convolutional feature map is flattened. 

We also ran experiments with another CNN architecture, where instead of flattening the final convolutional feature map,  global average pooling \cite{Srinivas} was performed to obtain fixed length feature maps. This solved the problem of large number of parameters in the fully connected layers. An added advantage of this architecture is that it can handle variable sized inputs, which is typical in speech data. The global average pooling layer allowed us to increase the size of filters in the initial convolutional layers and this led to better performance. The champion model on 14 class prediction was obtained with $12\times12$ filters and $7\times7$ filters in the first and second layers respectively.

Gaussian HMM and GMM-HMM were implemented with MFCCs and Log-melspectrogram features. HMMs with MFCCs gave around 20 \% accuracy and the log-mel spectrogram models gave 32 \% accuracy. This is probably due to the poor representative power of HMMs in general. Models such as 3D-CNNs, CNN-LSTMs and LSTMs were also trained with MFCC and Log-Mel spectrogram inputs for comparison of their performance in our task.
\begin{figure}[htp]
    \centering
    \includegraphics[width=13.5cm]{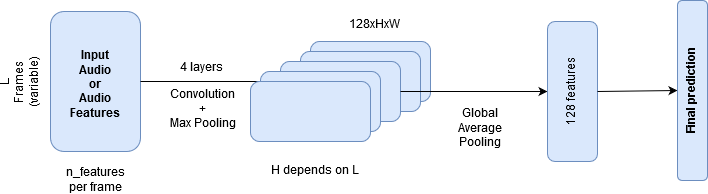}
    \caption{Visual representation of 2D CNN with Global average pooling layer }
    \label{fig:validation}
\end{figure}
\\
In order to have a deployable model, we wanted to ensure that our model can atleast broadly classify emotions either as positive or negative. Hence, we developed a 4 layer-2D CNN (with global average pooling) model to predict if the speaker's emotion is positive (neutral, happy, surprised) or negative (angry, sad, fearful, disgust). The model was trained on log-mel spectrogram features and the resulting model was able to make this binary classification with 88\% accuracy on unseen test and validation data. Given that the models together perform well in both broadly classifying the emotions and identifying the specific emotion, this model can be readily deployed.
\section{Results and Conclusions}
We have conducted an in-depth analysis of different feature engineering and modelling methods for emotion recognition. We obtain much better results with engineered features such as MFCCs and Log-mel spectrogram than the raw audio input, which is most likely due to the shortage of data. Although, MFCCs are the widely used features for speech based emotion recognition, we find that the Log-Mel spectrogram features are conclusively better at this task. We observe that gender-specific emotion classification leads to higher performance. This is due to the pitch and energy differences in the average male and average female voice, which makes patterns in male emotions different from female emotions. Also adding features such as pitch and energy to the MFCCs improved the model performances, which implies that the MFCC featues lack pitch and energy information, which is critical in emotion prediction. We also observe that 2D CNN gives the best performance over models such as 3D CNNs, LSTMS, CNN-LSTMs. We also tested our models on the TESS dataset and the 7 class prediction model achieved an accuracy of 62\%, where the chance accuracy (proportion of majority class) in TESS is 14\%. Therefore the RAVDESS-trained models are robust and have not overfit to the two sentences in the RAVDESS dataset. As mentioned earlier, the lower accuracies can be explained by the subjective nature of emotion perception by humans, which significantly complicates our problem.

Below is a summary of best performing models for different model architectures and input features.

\newcounter{rownumbers}
\newcommand\rownumber{\stepcounter{rownumbers}\arabic{rownumbers}}

\begin{center}

\scalebox{0.8}{
\begin{tabular}{|c|c|c|c|c|} 
\hline
 \textbf{Features} & \textbf{Architecture} &\textbf{No. of Class} & \textbf{Validation Acc.} & \textbf{Test Acc.} \\ \hline
 Pure Audio & 1D CNN + LSTM  & 12 & 61.6\% & 48.8\%\\ \hline
 Log Mel Spectogram & 4 Layer 2D CNN  & 12& 70.31\% & 65\%\\ \hline
 29 Coefficients:MFCC+Delta & 1 Layer 2D CNN & 12 & 56\% & 53\% \\ \hline
 Log Mel Spectogram & HMM & 12& - & 31.25\%\\ \hline
 Log Mel Spectogram & 3 Layer 3D CNN & 12& 66\% & 55\%\\ \hline
 \textbf{Log Mel Spectogram} & \textbf{2D CNN with Global Avg. Pool} & \textbf{14}& \textbf{70}\% & \textbf{66}\%\\ \hline
 \textbf{Log Mel Spectogram} & \textbf{2D CNN with Global Avg. Pool} & \textbf{2}& \textbf{90}\% & \textbf{86}\%\\ \hline
\end{tabular}}
\end{center}

The performance of the champion model (4 layer 2D CNN model with global average pooling on log-mel spectogram features) on 14 class prediction is shown below.
\footnotesize{
\begin{center}
\hspace*{0cm}\begin{tabular}{|c|c|c|} 
\hline
\textbf{Metric} & \textbf{Validation Set} &\textbf{Test Set}\\ 
\hline
Top-3 Categorical Accuracy & 90.8\% & 91.38\%\\
\hline
Top-2 Categorical Accuracy & 84.6\% & 84.13\%\\
\hline
Accuracy & 70\% & 66\%\\
\hline
F1-Score & 69.37\% & 63.6\%\\\hline
\end{tabular}
\end{center}
}
\begin{figure}[htp]
    \centering
    \includegraphics[width=10cm]{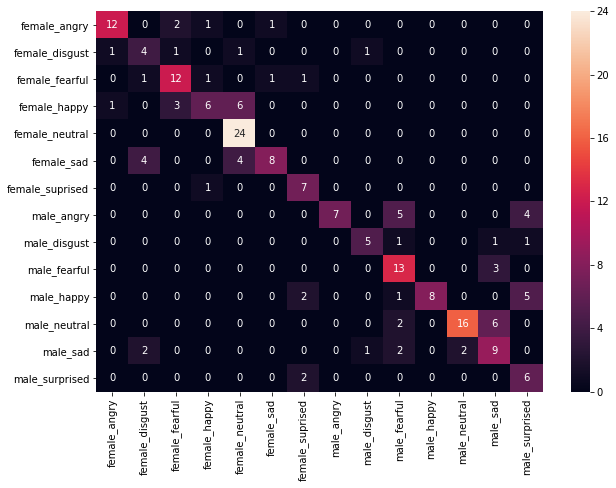}
    \caption{Confusion matrix of Champion Model}
    \label{fig:cm_14}
\end{figure}

When we tested our best model in the TESS dataset, it was observed that the model was misclassifying the gender as male for one of the female speakers. When we looked at the data, we found that the lady was 64 years old and her voice had lower pitch than most young and middle-aged women, which made her sound like an average male. Since voice features are strongly dependent on the gender and age group of a speaker, we should use a hierarchical structure, where multiple models are trained for different genders and different age groups for improvement in performance. In this structure, the age group and gender of a speaker would be predicted first and based on these, a model would be chosen for further classification. We can also visualise the regions of the input that maximally activate correct neurons in the final layer. This could give us some insight about what exactly the CNN is looking at in the log-mel spectrogram to discriminate between features. Understanding this could help us come up with novel feature engineering methods that work well. We think that the patterns corresponding to various emotions occur in varying widths across the input data (Log-mel spectrogram). So, Atrous spatial pyramid pooling (ASPP) \citep{Chen}  can also be applied, because learning features from multiple fields of view (larger and smaller) in this case would enable the CNNs to learn better features. Similarly, attention mechanisms can also be implemented in future works.

\newpage
\section{Appendix}
\subsection{Model Comparison Table }

\begin{center}
\scalebox{0.78}{
\hspace*{0cm}\begin{tabular}{|c|c|c|c|c|c|} 
\hline
\textbf{Features} & \textbf{Architecture} &\textbf{No. of Class} & \textbf{Validation Acc.} & \textbf{Test Acc.} \\ 
\hline
 Pure Audio & 1D CNN + LSTM & 12 & 61.6\% & 48.8\%\\
\hline
 Pure Audio & 2 Layer 1D CNN  & 12 & 58.5\% & 46.6\%\\
\hline
Log Mel Spectrogram & HMM&  12 & - & 31.25\%\\
\hline
 Pure Audio & HMM& 6 & - & 23.5\%\\
\hline
 13 MFCC Coefficients & HMM& 12 & - & 28.2\%\\
\hline

 Log Mel Spectogram & 2 Layer 2D CNN  & 12& 59\% & 54 \%\\
\hline
 Log Mel Spectogram & 3 Layer 2D CNN  & 12& 67.2\% & 61 \%\\
\hline
 Log Mel Spectogram & 4 Layer 2D CNN  & 12& 71.35\% & 65 \%\\
\hline
 Log Mel Spectogram & 4 Layer 2D CNN+LSTM  & 12& 63.4\% & 43 \%\\
\hline
 Log Mel Spectogram & 5 Layer 2D CNN  & 12& 70.83\% & 64 \%\\
\hline
 Log Mel Spectogram & 6 Layer 2D CNN  & 12& 76.01\% & 63 \%\\
\hline

 29 MFCC+Delta Coefficients & 1D CNN with Avg. Pooling & 6 & 61\% & 50\%\\
\hline
 Pitch, Magnitude, RMSE and its deltas & 3 Layer 2D CNN with Avg. Pooling & 6 & 51\% & 50\%\\
\hline
 29 MFCC+Delta+DeltaDelta Coefficients & 3 Layer 2D CNN with Avg. Pooling  & 6 & 42\% & 47\%\\
\hline
 29 MFCC+Delta Coefficients & 1 Layer 2CNN with Avg. Pooling & 6 & 56\% & 53\%\\
\hline
 12 MFCC Coefficients & GMMHMM& 12 & - & 21\%\\
\hline
 Log Mel Spectogram & 3 Layer 3D CNN  & 12& 66\% & 55 \%\\
\hline
 40 MFCC Coefficients + Deltas & 3 Layer DNN  & 6 & 51\% & 55 \%\\
\hline
 25 MFCC Coefficients & 3 Layer 3D CNN  & 12& 59\% & 53 \%\\
\hline
 Log Mel Spectogram &  256 Units LSTM  & 12& 63\% & 50 \%\\
\hline
 25 MFCC Coefficients &  256 Units LSTM  & 12& 53\% & 54 \%\\
\hline
 Log Mel Spectogram & 2D CNN With Avg. Pooling & 14& 70\% & 66 \%\\
\hline
 Log Mel Spectogram & 2D CNN With Avg. Pooling  & 4& 90\% & 86 \%\\
\hline

\end{tabular}}
\end{center}
\newpage
\subsection{Mel Spectrum Cepstral Coefficients (MFCC)}
The below figure shows the MFCC, Delta and Delta-Delta coefficient plots for a male neutral emotion audio.
\begin{figure}[htp]
    \centering
    \includegraphics[width=13.5cm]{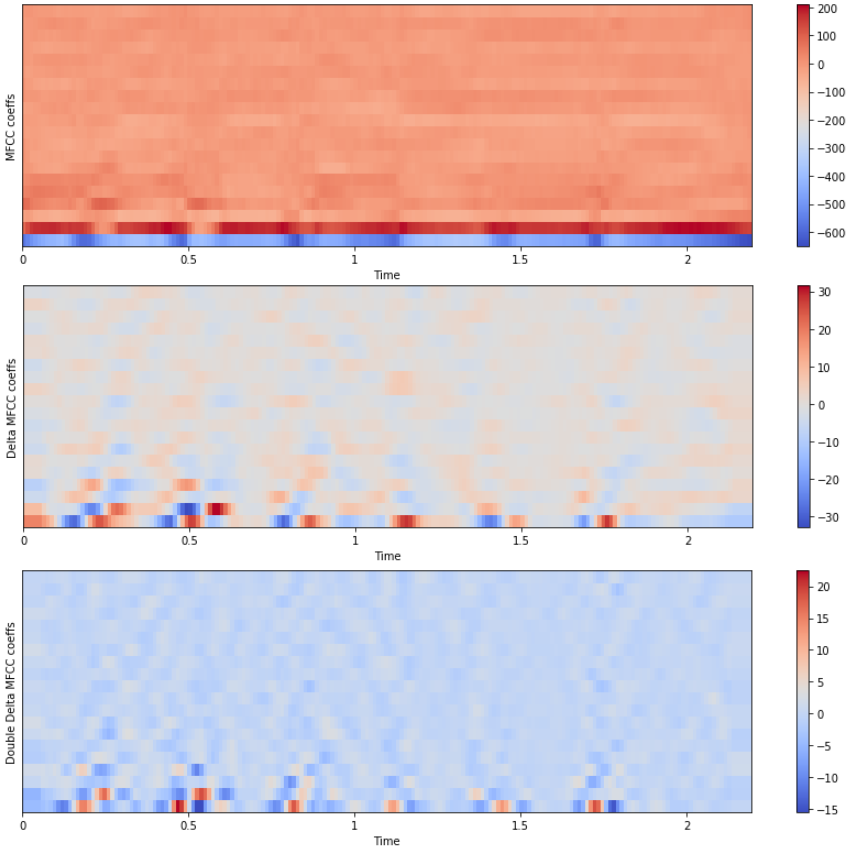}
\caption{MFCC, Delta and Delta-Delta Coefficients}
\end{figure}

The below figure shows the comparison of MFCC Delta coefficients of the audios enacted by male and female actor for the statement "Kids are talking by the door" with angry emotion
\begin{figure}[htp]
    \centering
    \includegraphics[width=13.5cm]{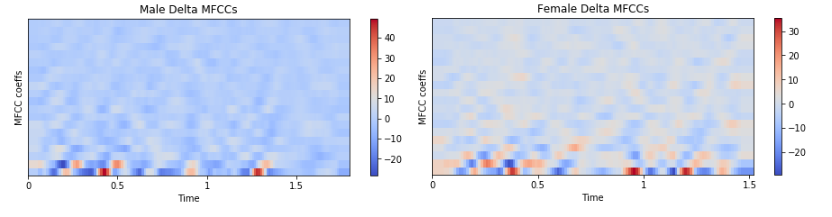}
\caption{Delta and Delta-Delta Coefficients}
\end{figure}

\newpage

\subsection{Chroma Features}
Chroma features are representation for audio in which the entire spectrum is projected onto 12 bins representing the 12 distinct semitones (or chroma) of the musical octave\cite{Muller}. Since RAVDESS dataset also has songs, we conducted initial analysis using Chroma. On a overall scale, other features exhibit clear patterns as compared to Chroma Spectogram. 

\begin{center}
\begin{figure}[htp]
    \centering
    \includegraphics[width=13.5cm]{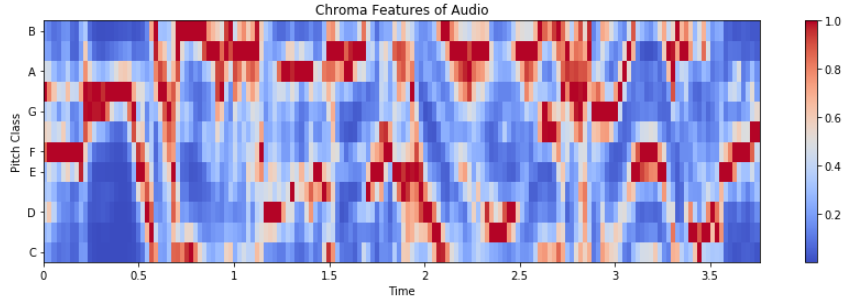}
\end{figure}
\end{center}

\subsection{Zero-crossing rate}
The zero-crossing rate is the rate of sign-changes along a signal, i.e., the rate at which the signal changes from positive to zero to negative or from negative to zero to positive\cite{Chen}.

\begin{center}
\begin{figure}[htp]
    \centering
    \includegraphics[width=13.5cm]{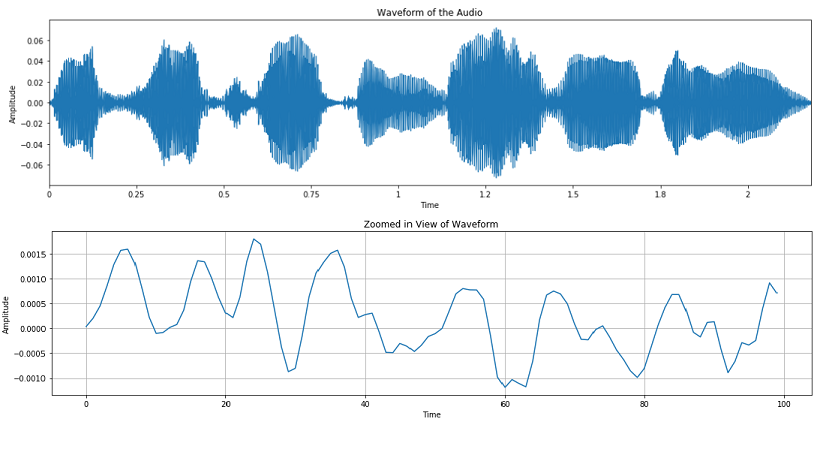}
\end{figure}
\end{center}
\newpage

\subsection{Log Mel Spectogram }
The below figure shows the Log Mel Spectogram plots for a Male Actor audios enacting the statement "Kids are talking by the door" with different emotions.

\begin{figure}[htp]
    \centering
    \includegraphics[width=13.5cm]{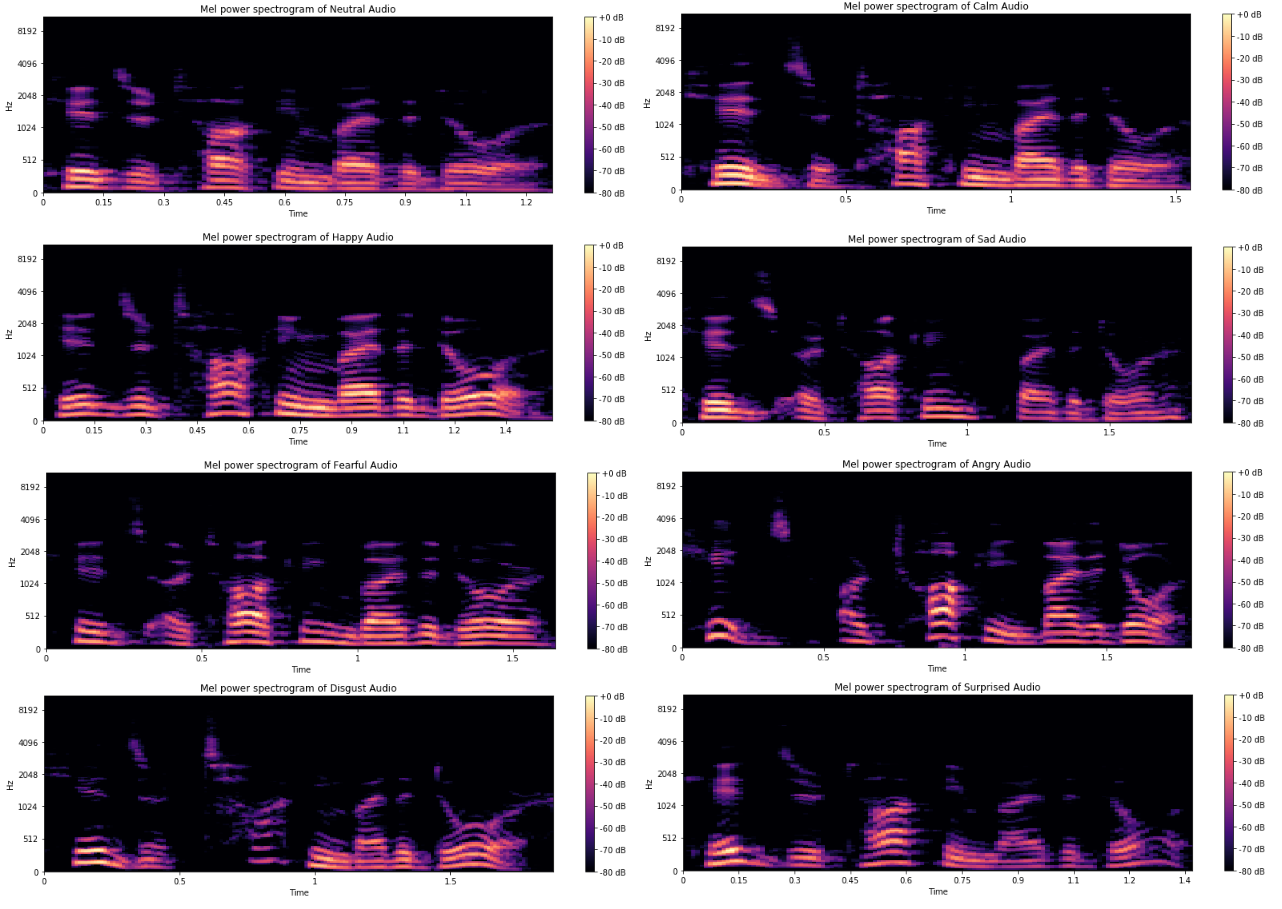}
    \caption{Log-Mel Spectograms of Audios with different emotions}
\end{figure}

The below figure shows the comparison of Log-Mel Spectograms of the audio enacted by male and female actor for the statement "Kids are talking by the door" with angry emotion.
\begin{figure}[htp]
    \centering
    \includegraphics[width=13.5cm]{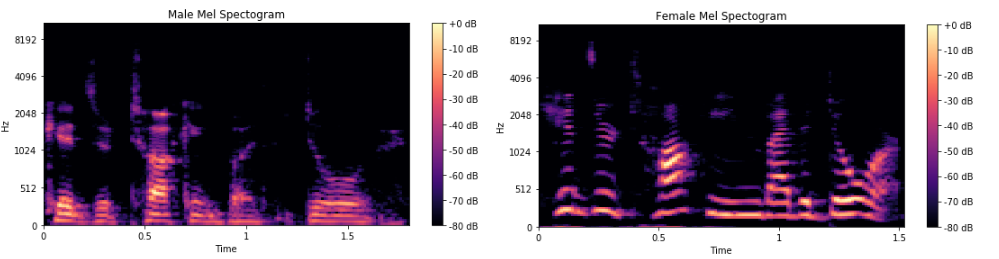}
\caption{Delta and Delta-Delta Coefficients}
\end{figure}

\end{document}